\documentstyle[12pt]{article}
\font\tenbf=cmbx10
\font\tenrm=cmr10
\font\tenit=cmti10
\font\elevenbf=cmbx10 scaled\magstep 1
\font\elevenrm=cmr10 scaled\magstep 1
\font\elevenit=cmti10 scaled\magstep 1

\renewenvironment{thebibliography}[1]
 { \elevenrm
   \begin{list}{\arabic{enumi}.}
    {\usecounter{enumi} \setlength{\parsep}{0pt}
     \setlength{\itemsep}{3pt} \settowidth{\labelwidth}{#1.}
     \sloppy
    }}{\end{list}}

\begin{document}
\begin{center}{{\tenbf QUANTUM MECHANICS OF THE ELECTRIC CHARGE \\ }
\vglue 1.0cm
{\tenrm A. STARUSZKIEWICZ \\}
\baselineskip=13pt
{\tenit Institute of Physics, Jagellonian University, Reymonta 4 \\}
\baselineskip=12pt
{\tenit 30 059 Krak\'ow, Poland \\}
\vglue 0.8cm
{\tenrm ABSTRACT}}
\end{center}
\vglue 0.3cm
{\rightskip=3pc
 \leftskip=3pc
 \tenrm\baselineskip=12pt
 \noindent
A simple argument against the existence of magnetic monopoles is given.
The argument is an important part of the quantum theory of the electric charge
developed by the author.
\vglue.8cm
``The same modification of the (Maxwell --Lorentz) theory which contains
$e$ as a consequence, will also have the quantum structure of radiation
as a consequence.''

{\tenit  Albert Einstein}

({\tenit Phys. Zeit.} {\tenbf 10 } (1909) 192)
\vglue 0.6cm}
{\elevenbf\noindent 1.Introduction}
\vglue 0.4cm
\baselineskip=14pt
\elevenrm
This paper is dedicated to Professor ~Yakir Aharonov on the occasion of his
$60^{\rm th}$ birthday.  The subject of the paper, quantum mechanics of the
 electric
charge, is based on the notion of {\em phase}, this elusive concept
which has always fascinated Professor Aharonov.

The electric charge $Q$ and the phase $S(x)$ of a (second quantized)
charged system are canonically conjugated variables:
\begin{equation}
  \left[ Q , S(x) \right] = ie, \hspace{1em} (\hbar = 1 = c)
\end{equation}
$e$ being the elementary charge. Proof of this theorem  is given in
${}^1$.
Here I will make only two rather obvious comments.

Eq.(1) does explain quantization of the electric charge $Q$ in units equal
 to the constant $e$:
\[ Q = ne, \hspace{1em} n=0, \pm 1, \pm 2, \dots .\]
It does not, however, explain the universality of the electric charge i.e.
the fact that e.g. the electric charge of the electron seems to be mathematically
 equal to the electric charge of the proton. Indeed, since the constant $e$
in Eq.(1) is arbitrary, we cannot ~exclude theoretically a situation in which
$e=e_1$ for one charged system and $e=e_2 \not= e_1$ for another system.
\vglue 0.6cm

{\elevenbf\noindent 2. The phase $S(x)$ can be uniquely determined at the
spatial infinity}
\vglue 0.4cm
$x$ in Eq.(1) is an arbitrary spatio-temporal point.
 Let us imagine that $x$ tends to the spatial infinity:
\[ xx \equiv (x^0)^2 - (x^1)^2 -(x^2)^2 -(x^3)^2 \rightarrow -\infty. \]
Mathematically-minded readers will object that we are not allowed
to fix, even in the form of a limit, the argument of an
operator-valued distribution. True. The argument which follows
is physical rather than mathematical, it constitutes a piece of
theoretical rather than mathematical physics.

At the spatial infinity there is
 only one function which can
possibly play the role of phase. This function must be equal to
\begin{equation}
 S(x) = - e x^{\mu} A_{\mu}(x),
\end{equation}
where $e$ is a constant proportionality factor and $A_{\mu}(x)$
is the electromagnetic potential. To see this one has to note
that at the spatial infinity the electromagnetic field is free,
\[ \partial^{\mu} F_{\mu \nu} \equiv 4 \pi j_{\nu} = 0 \]
and homogeneous of degree $-2$, $F_{\mu \nu}(\lambda x)=
\lambda^{-2} F_{\mu \nu}(x)$ for each $\lambda > 0$ ${}^2$.
The field is free because the electric current $j_{\nu}$, being
carried by massive particles, must be confined to the future and
past light cone. It must be homogeneous of degree $-2$ because,
as seen e.g. in the static case, the charge generated monopole term
dominates dipole and higher terms.

Consider a classical electromagnetic field which is free and
homogeneous of degree $-2$; assume that its potential is
homogeneous of degree $-1$, which is natural. Let us form two
vectors, \\
\[  F_{\mu \nu}(x) \, x^{\nu}  \hspace{1em} {\rm and }
\hspace{1em}  \frac{1}{2} \epsilon^{\mu \nu \rho \sigma} x_{\nu}
F_{\rho \sigma}(x), \]
where $x$ is the radius vector in the Lorentzian reference frame
in which the homogeneity condition holds.

 The two vectors given
above determine the tensor $F_{\mu \nu} $ in a purely algebraic
way. Both these vectors are gradients of homogeneous of degree
zero functions:
\[ F_{\mu \nu}(x) \, x^{\nu} = \partial_{\mu} e(x),
\hspace{1em} \frac{1}{2} \epsilon^{\mu \nu \rho \sigma} x_{\nu} F_{\rho
\sigma}(x) = \partial^{\mu} m(x).\]
$e(x)$ and $m(x)$ denote ``electric'' and ``magnetic'' parts
respectively. $e(x)$ can be easily calculated:
\begin{eqnarray*}
\lefteqn{ F_{\mu \nu}(x) \, x^\nu = \left[ \partial_\mu A_\nu(x) -
\partial_\nu A_\mu(x) \right] x^\nu =} \\
& & \partial_\mu\left[A_\nu(x) \, x^\nu\right] -\delta^\nu_\mu A_\nu(x)
- x^\nu \partial_\nu A_\mu(x) =
\partial_\mu\left[ x^\nu A_\nu(x) \right]
\end{eqnarray*}
because
\[ x^\nu \partial_\nu A_\mu(x) = - A_\mu(x) \]
from the Euler theorem on homogeneous functions.

I maintain that $m(x)$ must be a constant. This is an argument
against the existence of magnetic monopoles which, to the best
of my knowledge, has never been put forward before. (The argument
given by Dr. Herdegen ${}^3$ is different.)

To see this let us calculate the  Lagrangian density
\begin{equation}
dx^0 dx^1 dx^2 dx^3 F_{\mu \nu} F^{\mu \nu}
\end{equation}
for a homogeneous of degree $-2$ field $F_{\mu \nu}$, using the
spherical coordinates
\begin{eqnarray*}
 x^0 & = & \xi^0 \sinh \xi^1, \\
 x^1 & = & \xi^0 \cosh \xi^1 \sin \xi^2 \cos \xi^3, \\
 x^2 & = & \xi^0 \cosh \xi^1 \sin \xi^2 \sin \xi^3, \\
 x^3 & = & \xi^0 \cosh \xi^1 \cos \xi^2,
\end{eqnarray*}
\[ 0<\xi^0 <\infty, \hspace{1em} -\infty <\xi^1 <+ \infty, \hspace{1em}
 0\leq \xi^2\leq \pi, \hspace{1em} 0\leq \xi^3 <2 \pi. \]
These coordinates cover in an obvious way the spatial infinity
we are interested in. Note that $\xi^0$ is a space-like
coordinate while $\xi^1$ is a time-like coordinate. A simple
calculation gives
\[ dx^0 dx^1 dx^2 dx^3 F_{\mu \nu} F^{\mu \nu} = 2 \frac{d\xi^0}{\xi^0}
\sqrt{g} \, d\xi^1 d\xi^2 d\xi^3 \left(- g^{ik}
\partial_i e \, \partial_k e + g^{ik} \partial_i m \, \partial_k m\right).\]
Here
\[ g_{ik} =(\xi^0)^{-2} g_{\mu \nu} \frac{\partial x^\mu}{\partial \xi^i}
\frac{\partial x^\nu}{\partial \xi^k}, \hspace{1em} i,k=1,2,3, \]
is the metric on the spatial infinity.

The Lagrangian density ~(3) is seen to be a difference of two
identical Lagrangian densities. Thus only one of them can have
the correct sign i.e. the sign which, upon quantization, would
give a positive definite inner product. The part with the right
sign is {\em called} electric, the part with the wrong sign is {\em called}
magnetic and must be put equal to zero.

Now, the Gauss theorem says that the total charge $Q$ is
determined by the   electromagnetic field at the spatial
infinity. In the quantum theory the charge operator $Q$ must have
its  canonically conjugated variable $S(x)$. Thus $S(x)$ must
have a ``tail'' which does not vanish even at the spatial
infinity. We have seen, however, that there is exactly one
function, namely $x^\mu A_\mu (x)$, which can play the role of
the ``tail''. Hence, there must exist a constant $e$ such that
at the spatial infinity
\setcounter{equation}{1}
\begin{equation}
 S(x) = - e x^\mu A_\mu (x).
\end{equation}
The constant $e$ in this equation is identical with the constant
$e$ in Eq.(1). This is a hypothesis substantiated in the next
section.
\vglue 0.6cm

{\elevenbf \noindent 3. The proportionality factor in the phase}
\vglue 0.4cm

The two equations \\
\vspace{-2ex}
\begin{center} \begin{tabular}{|c|}
                            \hline
                                {} \\
  $  \left[Q, S(x)\right] = ie,  $ \\
                                {} \\
   $  S(x) = -e x^\mu A_\mu (x), $ \\
                                {} \\
              \hline
        \end{tabular}
\end{center}
\vspace{2ex}
constitute together a closed theory, the quantum mechanics of the
electric charge. It is important to understand correctly the
epistemological status of both equations. The first equation is
simply a theorem in the Q.E.D. which, by continuity, is assumed
to hold also at the spatial infinity. The second equation is a
hypothesis; one can give several arguments supporting Eq.(2) but
all those arguments do not amount to a proof. Here are two simple
arguments, to be added to those which I have given
elsewhere ${}^1$.

Take the Coulomb field of the charge $Q$ at rest:
\[ A_0 = \frac{Q}{r}, \hspace{1em} A_1 = A_2 = A_3 = 0. \]
Its phase, according to Eq.(2), is
\[ S(x) = - e \frac{Q}{r} t = - e Q \frac{t}{r}. \]
During the eternity of time available at the spatial infinity,
\[ -r < t < r, \]
the phase $S(x)$ changes from $eQ$ to $-eQ$. Take now the
hydrogen atom with the ~nuclear charge $Q$ and the electron
charge $e$ and assume that the radius of its circular orbit tends
to infinity. During the eternity of time available,
\[ -r < t < r , \]
the electromagnetic phase of the electron wave function,
\[ -e \int \! A_\mu (x) \, dx^\mu , \]
will change by the same amount:
\[- e \int_{-r}^r \frac{Q}{r} \, dt = - 2 e Q. \]
Thus the phase given by Eq.(2) changes as the true phase of the
electron wave function in an infinitely large hydrogen atom.

The phase of the Coulomb field ,
\[ S(x) = - \frac{eQ}{r} t \]
may be compared with the phase of the wave function of a
stationary state, $-Et$, $E$ being the energy of the stationary
state. Thus $S(x)$ looks like the phase of a stationary state
driven by the Coulomb energy $eQ/r$. Again, this is not a proof
but a heuristic argument supporting Eq.(2).

Equations (1) and (2) together do allow to explain the universality
of the electric charge. To be  more precise, they allow to prove
the following theorem: the total charge of the universe is always
a multiple of a single constant. To apply this to the electron
or to the proton one must be able to estimate the accuracy with
which, under specific observational circumstances, they can be
considered as isolated universes. The experimental equality of
electron's and proton's charge shows that this accuracy is indeed
extremely high.
\vglue 0.6cm

{\elevenbf \noindent 4. Acknowledgement~}
\vglue 0.4cm

I thank Professor Pawel O. Mazur for encouragement and several
useful discussions.
\vglue 0.6cm

{\elevenbf\noindent 5. References \hfil}
\vglue 0.4cm

\vglue 0.5cm
\end{document}